\documentclass[9pt,twocolumn,twoside]{osajnl}
\journal{ol}
\usepackage[utf8]{inputenc}\setboolean{shortarticle}{true}
\setboolean{shortarticle}{true}
\usepackage{hyperref}
\usepackage{graphics}
\usepackage{graphicx}
\usepackage{amsmath}
\usepackage{lipsum}
\usepackage{mathtools}
\usepackage{cuted}
\title{Observation of geometric phase for unpolarized and partially polarized light fields}

\author[1*]{Bhaskar Kanseri}
\author[1]{Rohit Gupta}

\affil[1]{Experimental Quantum Interferometry and Polarization, Department of Physics, Indian Institute of Technology Delhi, Hauz Khas, New Delhi-110016, India}
\affil[*]{Corresponding author: bkanseri@physics.iitd.ac.in}

\dates{Compiled \today}
\ociscodes{(260.5430) polarization; (030.0030) Coherence and statistical optics.}
 
\doi{\url{http://dx.doi.org/10.1364/ol.XX.XXXXXX}}

\begin{abstract}
Geometric phase, owing to its topological nature and properties of fault tolerance, plays an important role in devising real world applications in both classical and quantum domain. For classical systems, geometric phase has been observed and studied so far for fully polarized light only. Using an interferometric experiment we demonstrate, for the first time, the existence of Pancharatnam-Berry phase for states covering all empty space inside the Poincar\'e sphere namely the unpolarized and partially polarized light fields. The observed geometric phase is found identical to its fully polarized counterpart in excellent agreement with the theoretical predictions.
\end {abstract}

\begin{document}

\maketitle

The Pancharatnam--Berry phase,  first formulated in 1950's by Pancharatnam in classical optics \cite{panch}, and later by Berry for dynamical quantum systems \cite{berry84, berry87}, has been demonstrated both theoretically and experimentally for polarized light fields when the state of polarization (SOP) traces out a closed loop on the surface of Poincar\'e sphere \cite{panch, berry84, berry87, bhandari88, simon88, martinelli90, poin92}. The phase acquired owing to the geometry of the curved polarization space is equal to half the solid angle subtended by the SOP trajectory, and is therefore also referred as `geometric phase' \cite{shapere89}, finding applications in wide areas of physics such as in optics, condensed matter physics, quantum physics, topological physics, atomic and molecular physics etc \cite{falci00, biener02, lee17, slussarenko16, jisha19, decamps17}. Despite of having such widely explored area for last seven decades, geometric phase is still associated with pure quantum states and fully polarized light i.e., with states having unit degree of polarization (DOP) only. Owing to tremendous applicability of geometric phase in development of devices \cite{bhandari97}, its implementation for all polarizations and for mixed quantum states may significantly accelerate its existing usage in cutting edge multidisciplinary areas of research including quantum computing \cite{jones00, leibfried03}, characterizing the evolution of several kinds of physical systems \cite{erik00, carollo03}, in fundamental studies \cite{leek07, pechal12}. \textcolor{black}{More recently, adiabatic geometric phase has been observed in non-linear optics for frequency conversion and three-wave mixing, which may lead in harnessing several new applications such as all-optically controlled geometric phase elements etc \cite{arie18, arie19, arie20}.}  

The polarization properties of any stationary light field can be expressed in terms of a unique coherency matrix $J$ \cite{wolf59} containing all four set of correlations between the electric field components, \textcolor{black}{as
\begin{equation}J=
\begin{bmatrix}
\langle E_p^*(\textbf{r},t)E_p(\textbf{r},t) \rangle & \langle E_p^*(\textbf{r},t)E_s(\textbf{r},t) \rangle \\ 
\langle E_s^*(\textbf{r},t)E_p(\textbf{r},t) \rangle & \langle E_s^*(\textbf{r},t)E_s(\textbf{r},t) \rangle
\end{bmatrix},
\end{equation}
where $\langle ..\rangle$ denotes the ensemble average \cite{wolf03}.  The principal diagonal elements of this matrix represent intensities corresponding to $p$ and $s$ components of the light field, respectively. The DOP $P$ of the light field which is given by the ratio of intensity of polarized  light ($I_{pol}$) to the total intensity of the field ($I_{tot}$) is expressed in terms of matrix $J$ as \cite{wolf59}
\begin{equation}
P=\frac{I_{pol}}{I_{tot}}=\sqrt{1-\frac{4 \,\text{det} \,J}{(\text{Tr}\,J)^2}},
\end{equation}
where det is determinant and Tr is trace of the matrix. From Eqs. (1) and (2), it is apparent that out of the four elements of matrix $J$, off-diagonal element $\langle E_p^*(\textbf{r},t)E_s(\textbf{r},t)\rangle$ containing correlation between orthogonal components $p$ and $s$ play an important role in quantification of DOP at a space-time point $(\textbf{r},t)$. The degree of coherence (DOC) corresponding to the off-diagonal element yields as \cite{wolf03}
\begin{equation}
|\gamma|=\left|\frac{\langle E_p^*(\textbf{r},t)E_s(\textbf{r},t)\rangle}{\sqrt{\langle E_p^*(\textbf{r},t)E_p(\textbf{r},t)\rangle}\sqrt{\langle E_s^*(\textbf{r},t)E_s(\textbf{r},t)\rangle}}\right|.\end{equation}
From Eqs. (2) and (3), one can see that the DOP of the field is given by the maximum value of the DOC ($|\gamma|$)  between the orthogonal set of field components for which the two intensities are equal, i.e.$\langle |E_p(\textbf{r},t)|^2\rangle=\langle |E_s(\textbf{r},t)|^2\rangle$  \cite{wolf59}. Highest value of this correlation ($|\gamma|$=1) refers to fully polarized light (DOP=1) whereas null ($|\gamma|$=0) and intermediate ($0<|\gamma|<1$) values of the correlation refer to unpolarized (DOP=0) and partially polarized light ($0<\text{DOP}<1$), respectively}. Only fully polarized light  having different SOPs can be represented by points on the surface of a unit-radius sphere called the Poincar\'e sphere \cite{goldstein, wolf03}. A generalization, namely the three-dimensional (3D) Stokes space, is needed in which space inside the sphere represents partially polarized light fields and the origin of the sphere represents unpolarized light field. States having a constant DOP reside on a fixed radius spherical shell inside this 3D Stokes space.

In this letter, we report the existence of geometric phase for all kinds of polarization states of light including partially polarized and unpolarized light- a context for which the geometric phase has never been considered before. The experimental scheme consists of a source of tunable DOPs capable of producing arbitrarily polarized, partially polarized and near unpolarized light beam; along with a stable interferometric assembly, in which a set of quarter and half wave retarders are placed to make transformations on any spherical shell inside the 3D Stokes space. The experimentally measured topological phase is found to be identical for all incident DOPs of the light field, consistent with the simulated theoretical predictions.

In the classic work of Pancharatnam for fully polarized light, the interfering beams were essentially scalar fields \cite{panch}. In an intuitive manner, it implies that one must expect similar interference laws once light beams having same DOPs interfere, i.e, the trajectories of interfering beams remain on a spherical shell inside the 3D Stokes space \cite{goldstein}. This requires a scheme in which both the orthogonal decomposed components $p$ and $s$ of the field follow trajectories obtained using polarization preserving unitary transformations before interference. \textcolor{black}{The geometric phase, which depends only on the geometry of curved polarization space, is expected to behave identically irrespective of the transformations made on any spherical shell}. This would lead to identical geometric phase for all the DOPs of the light field provided that the trajectories for $p$ and $s$ components are identical. In spherical coordinates $(r,\theta,\phi)$, the solid angle subtended by a curved polarization space at the origin of the 3D Stokes space is given by $\Omega=\int\int\sin\theta d\theta d\phi$. As a consequence the geometric phase is given by $\Omega/2$, which clearly does not depend on the distance of the spherical shell from the origin $r$. Thus as long as the solid angles are identical, the acquired geometric phases must be the same irrespective of the DOP of the input field. As an example, the field of view or solid angle of 4$\pi$ corresponds to tracing of a complete spherical shell in the 3D Stokes space independent to its radius.  

\begin{figure}[t]
\centering
\includegraphics[width=\linewidth]{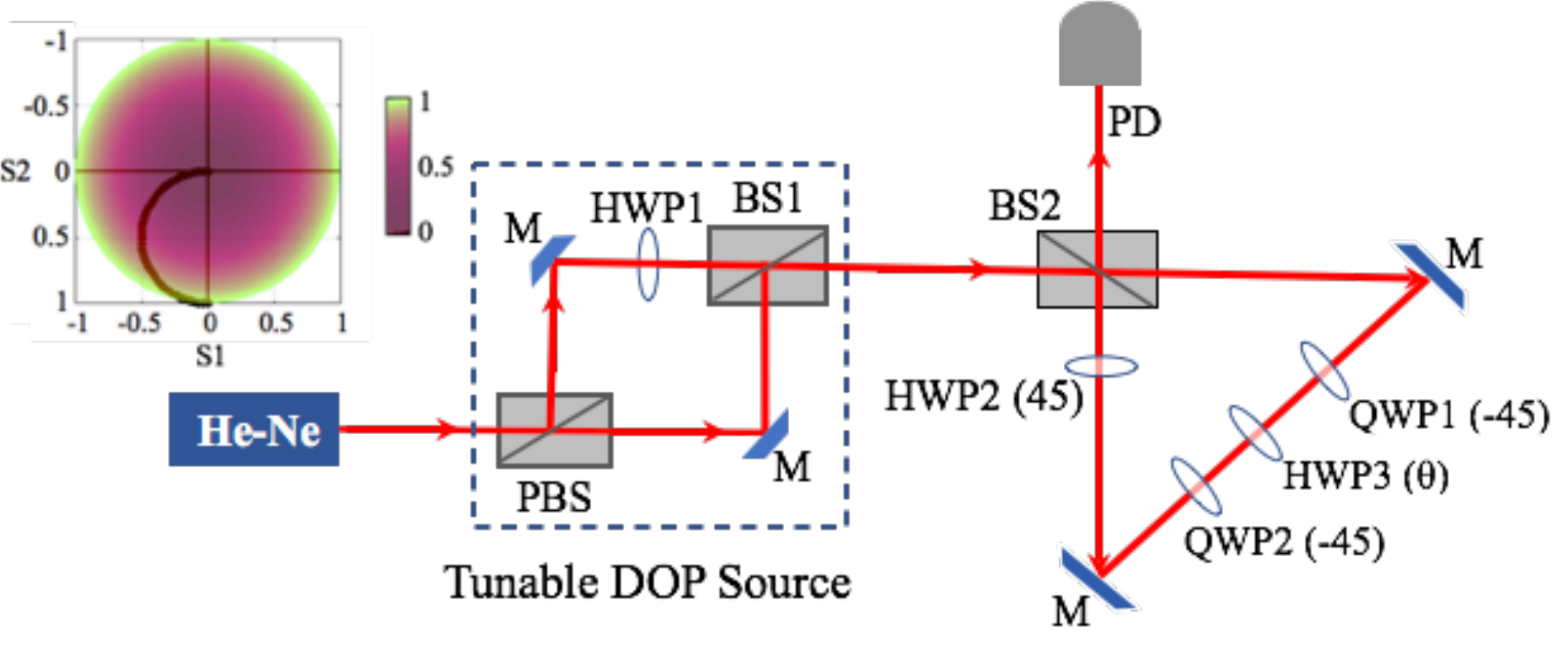}
\caption{Scheme of the experiment. A randomly polarized laser beam (He-Ne laser, 632.8nm) passes through a Mach-Zehnder assembly consisting of a polarizing beam splitter (PBS), a half-wave plate (HWP1), and a non-polarizing beam splitter (BS1) to make a tunable DOP source. The DOP of the output optical field from the first interferometer is measured using a Stokes measurement scheme consisting of a quarter-wave plate (QWP), a polarizer and a photodetector \cite{kanseri19}. The output beam with a known DOP is divided in two identical intensity beams using a non-polarizing beamsplitter (BS2). Both the beams traverse identical paths in a sagnac interferometer.  The polarization transformation is made using a phase-shifter consisting of a set of retardation plates QWP1, HWP3, and QWP2 with orientations as $-45^0$, $\theta^0$, $-45^0$; respectively. HWP2 having fast axis at $45^0$ with respect to horizontal polarization makes the output beam polarization orthogonal to the incident polarization. The beams interfere at BS2 and the interference fringe intensity is measured using a photodetector (PD). (Inset) Equatorial cross-section of 3D Stokes space showing DOP variation (in colour) from 0 to 1 of the tunable DOP source with associated Stokes parameters $S_1$ and $S_2$ (black curve) with $S_3=0$.}
\end{figure}

In order to observe geometric phase for all DOPs, first we construct a source of tunable DOP \cite{kanseri19}. As shown in Fig. 1, orthogonally polarized modes of a randomly polarized He-Ne laser beam are separated using a polarizing beam splitter and made to interfere again in a non-polarizing beam splitter after passing the reflected beam through a half wave plate (HWP1). \textcolor{black}{Since the laser is randomly polarized, there is null correlation between orthogonal field components having equal intensities, i.e. $\langle E_p^*(\textbf{r},t)E_s(\textbf{r},t)\rangle=0$, by selectively increasing intensity of one polarization component over the other, the output field could be changed from unpolarized to fully polarized. This is achieved in our case by rotating the angle of HWP1 from 0 to 45 degree, which gradually changes input $s$-polarization to $p$-polarization before mixing with the $p$ polarized light transmitted by PBS. A theoretical explanation of this behaviour can be found in \cite{kanseri19}.} Such configuration provides tunability in DOP with experimental values for our practical source changing from 0.03 (near unpolarized) to 0.99 (fully polarized) by changing the HWP1 angle from 0 to 45 degree obtained by measuring Stokes parameters using standard polarimetric assembly \cite{hauge76,kanseri19} . Since this source provides partial polarization by mixing \textcolor{black}{two uncorrelated linearly polarized light beams} (no circular polarization), resulting polarization states can be represented on the equatorial plane of the 3D Stokes space, as shown in the inset of Fig. 1.

A fixed DOP light beam coming from the tunable DOP source is further inserted in a Sagnac interferometer \cite{sagnac13} which houses a set of three birefringent plates (phase shifter) oriented in a manner (Fig.1) that each polarization component ($s$ or $p$) traces a closed loop of curved polarization space \cite{martinelli90} on a given spherical shell of 3D Stokes space. In the experiment, \textcolor{black}{since only unitary, energy conserving and polarization preserving transformations are made on the input light field using half-wave and quarter-wave plates placed on the path of the interferometer, the DOP of light remains preserved during these transformations, ensuring the trajectories remain on a spherical shell inside the Poincar\'e sphere.} Since in a Sagnac interferometer, both the interfering beams travelling in opposite directions and passing through birefringent (retarder) plates traverse exactly identical optical paths, the dynamical phase remains constant throughout the rotation of intermediate HWP3 \cite{simon88}, which infers that the measured phase is a topological geometric phase. Additionally, phase changes occurring due to mechanical, physical and environmental disturbances are nullified in this interferometric geometry. 

\begin{figure}[t]
\centering
\includegraphics[width=\linewidth]{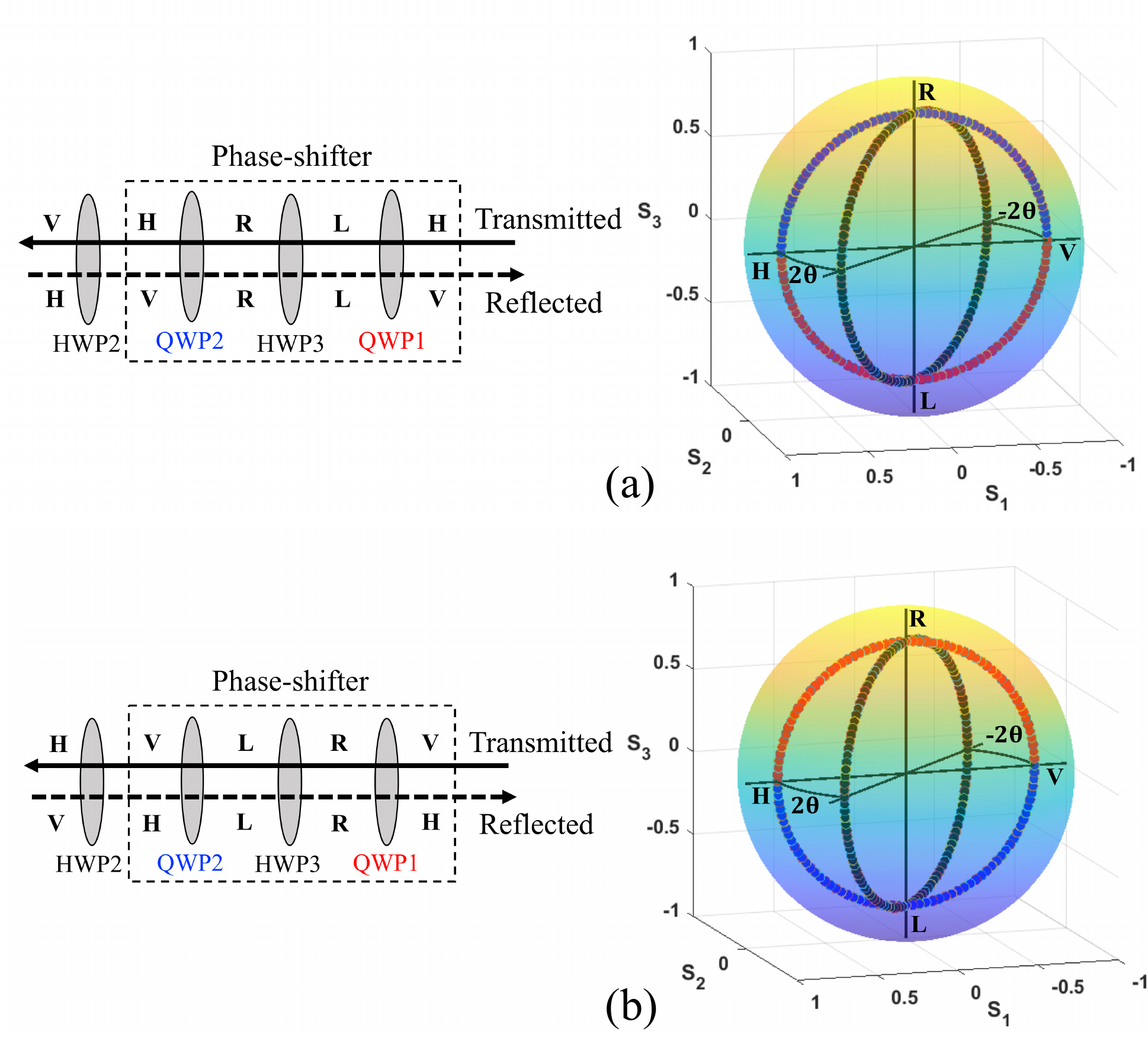}
\caption{ Trajectories of paths (right) traversed by polarization states on a spherical shell inside the coloured 3D Stokes space (corresponding to DOP 0.8) after passing through a set of birefringent plates of phase-shifter placed in Sagnac interferometer (left) for (a) $p$ polarized component, (b) $s$ polarized component. In both the cases the closed curves are identical implying same solid angles and hence same geometrically acquired phase shifts. Color of trajectory represents the transformation of polarization state by the corresponding birefringent plate in the phase-shifter.}
\label{fig:false-color}
\end{figure}

Fig. 2 demonstrates the simulated trajectories on a spherical shell of 3D Stokes space for $p$ and $s$ polarizations of input field for an arbitrary value of DOP (here 0.8) obtained using Mueller's matrix approach \cite{goldstein}. The reasoning behind this is as follows. The $p(s)$ component of the input beam transmitted by BS2 passes through quarter wave plate QWP1 having fast axis at $-45^0$ with respect to horizontal polarization, which converts the SOP to left (right) circularly polarized light keeping the DOP same. Further it passes through HWP3, with fast axis at angle $\theta$ with respect to QWP1, which converts SOP to right (left) circular polarization through an arc that cuts the equator of spherical shell at angle $2\theta$ away from the starting position \cite{martinelli90, goldstein}. Plate QWP2 closes the trajectory by bringing back it to the initial state $p(s)$ with path HLRH (VRLV) which is converted to $s(p)$ polarization using HWP2. \textcolor{black}{The solid angle $d\omega$ corresponding to any closed path on a spherical shell can be evaluated as \cite{goldstein}
\begin {equation}
d\omega= \frac{\text{dA}}{r^2},
\end{equation}
where $dA$ represents the area of the closed path and $r$ is the distance of the spherical shell from the origin of the sphere. In case of $2\theta$ angle, the area is given by ($2\theta\times 4\pi r^2)/{2\pi}$, which from Eq. (4) results to a solid angle of $4\theta$ for both $s$ and $p$ polarized components of light. The acquired geometric phase would then corresponds to half of the solid angle subtended to the origin of the sphere \cite{panch}.} Thus each of these closed curved polarization spaces subtends a solid angle of the $4\theta$ at the centre of the 3D Stokes space acquiring a phase shift of $2\theta$. Similarly for the $p(s)$ component of the light reflected by BS2 will follow identical paths VRLV (HLRH) in a complementary manner (see Fig. 2) exhibiting the same solid angle of $4\theta$ and acquired phase shift of $2\theta$. Using HWP2, the polarization state is made orthogonal to the incident one resulting to distinct observation of both the closed curved paths on the 3D Stokes space. Since both these trajectories are identical and overlapping each other, the total acquired phase shift for the combined ($s+p$) field would be half of the total solid angle, i.e. $4\theta$. Similar reasoning applies to any given DOP ($0\leq \text{DOP}\leq1$), for which both the traversed optical trajectories and hence corresponding solid angles would be identical in this scheme and the acquired phase shift would always be $4\theta$ only. Even for the unpolarized ($\text{DOP}=0$) state, for which no correlation (amplitude/phase) exists between $s$ and $p$ components of light (having intensity ratio 1), and is a point at the origin of 3D Stokes space, $p$ and $s$ components would undergo similar state transitions and the acquired phase shift would result the same value of $4\theta$. 

\begin{figure}[t]
\centering
\includegraphics[width=\linewidth]{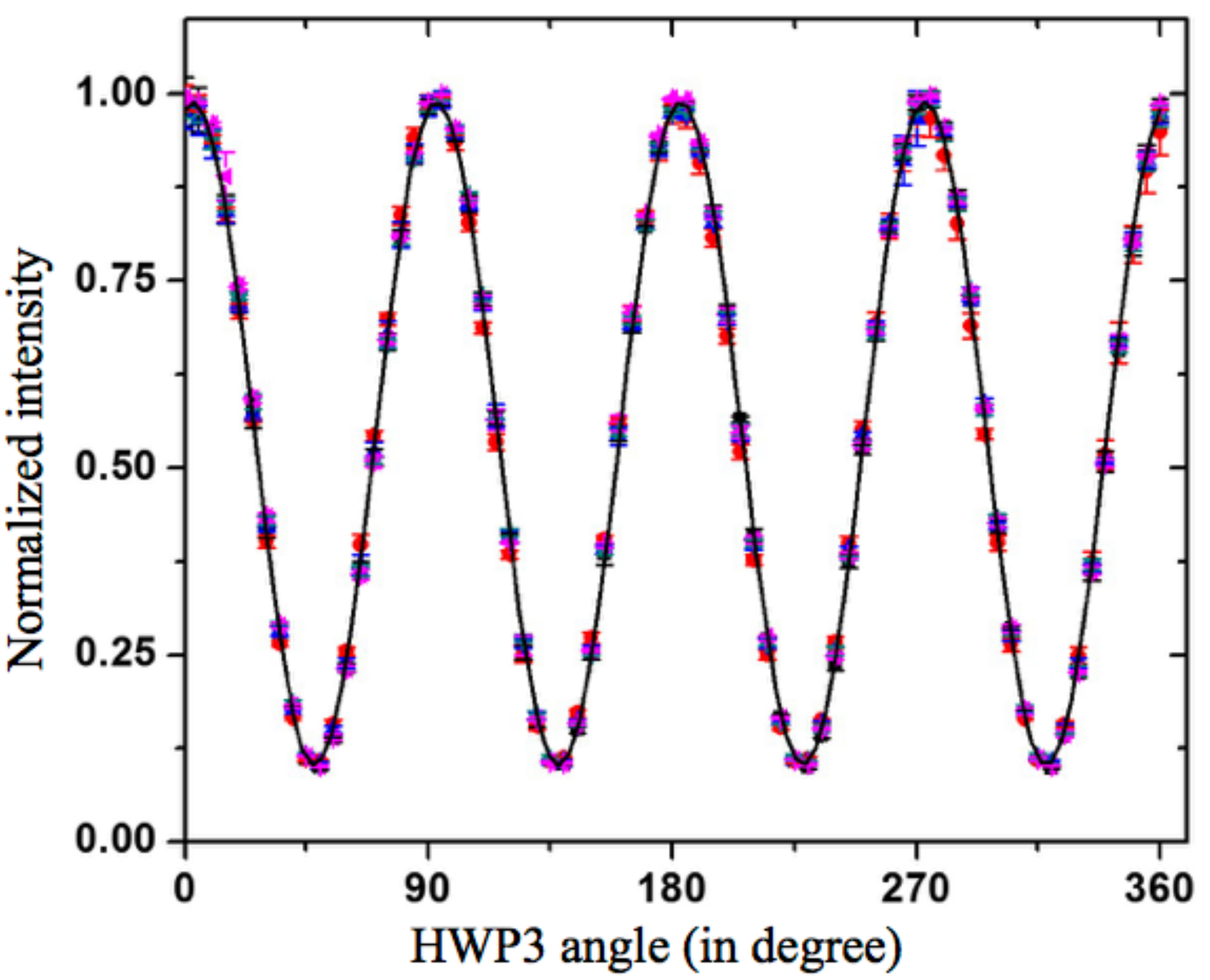}
\caption{Plot of normalized output intensity (measured by photodiode) with respect to the rotation angle $\theta$ of the HWP3 (Fig. 1) for different DOPs of the input light field. Notations for data points corresponding to DOPs: black square 0, red diamond 0.25, blue up triangle 0.5, green down triangle 0.75 and pink left triangle 1.  One can observe identical behaviour of intensity modulations for all kinds of input DOPs, \textcolor{black}{which results in overlapping of the different symbols making them non-separable}. The error bars show the uncertainty in the mean value of intensity calculated at a 95\% confidence level. Continuous black line is the theoretical fit corresponding to $4\theta$ angle using a cosine function. }
\label{fig:false-color}
\end{figure}

\textcolor{black}{For each setting of the HWP1 angle rotated in steps between 0 to 45 degree, which corresponds to change in DOP ranging from near 0 to near 1, HWP3 is rotated for a complete circle ($2\pi$ angle) to observe the effect of the geometric phase on the interference.} The normalized intensity variation at interference field as a function of the HWP3 angle measured using the photodiode is shown in Fig. 3, which manifests sinusoidal modulations.  A theoretical fit using cosine function of period $4\theta$ shows an excellent match with experimental data within the uncertainty of the measurement. This confirms that the acquired phase shift between $s$ and $p$ components of the field for a complete cycle of the HWP3 rotation is obtained as $4\theta$. This phase shift has been observed to be same for all DOPs of the input field ranging from fully polarized, partially polarized and unpolarized light fields (Fig. 3). Clearly, the geometric phase depends only on the geometry of the curved surface (solid angle circumvented by the polarization state path) and is observed to be independent of the DOC (correlation) between the orthogonal field components. This demonstrates the topological robustness of the geometric phase. In each of the cases, we obtain the interference visibility of nearly $87\%$ which is mainly limited due to the dissimilar transmission and reflection coefficients (asymmetry) of the beam splitter \cite{kanseri19} and imperfections of retarders employed in the interferometer. 

\textcolor{black}{An extension of Poincar\'e sphere for spirally polarized beams or vector vortex states namely the higher order Poincar\'e sphere \cite{alfano11} and more recently for fields propagating in inhomogeneous anisotropic media namely the hybrid-order Poincar\'e sphere \cite{fan15} have been proposed. These singular beams, which are essentially optical angular momentum states, find diverse applications in the realm of classical and quantum optics \cite{padgett09, padgett11}. Geometric phases were also demonstrated in these new kinds of Poincar\'e spheres showing a direct relationship with the angular momentum of light \cite{alfano11, fan15, dennis19}. It would be interesting to explore in the near future the geometric phase for optical states residing inside the higher-order and hybrid-order Poincar\'e spheres, in line with the study presented in this work, which could pave the way for several new developments and applications in this area.}

The striking connection between the Panchratnam phase (classical) with Berry phase (quantum) \cite{berry87} advocates similar behaviour of the quantum phase for unpolarized and partially polarized quantum states as they can be represented by states residing inside the Bloch sphere, which provides geometrical representation of quantum states analogues to Poincar\'e sphere for classical polarization states \cite{chuang04}. Mixed quantum states can be represented as non-unique convex sum of pure quantum states (density matrix representation), in similarity with partially polarized states which can be decomposed as sum of fully polarized states \cite{fano57, chuang04, wolf03}. Current study made for partially polarized and unpolarized states is expected to stimulate further experimental efforts in order to observe Panchrantam-Berry phase for mixed quantum states.

\section*{Funding Information}
Science and Engineering Research Board (SERB), India. 

\section*{Disclosures}
The authors declare no conflicts of interest.

\end{document}